\documentclass[a4paper]{article}
\usepackage[utf8]{inputenc}
\usepackage{setspace}
\usepackage{fullpage}
\usepackage{epsfig}
\usepackage{amsfonts}
\usepackage{amssymb}
\usepackage{amsmath}
\usepackage{subfig}
\usepackage{comment}
\usepackage{multirow}
\usepackage{soul}
\usepackage[dvipsnames]{xcolor}

\begin{document}

\title{Target Wake Time: Scheduled access in IEEE 802.11ax WLANs}
\author{Maddalena Nurchis, Boris Bellalta \\ Universitat Pompeu Fabra, Barcelona \\ \textit{name.surname@upf.edu}}

\date{}

\maketitle

\begin{abstract}
The increasing interest for ubiquitous networking, and the tremendous popularity gained by IEEE 802.11 Wireless Local Area Networks (WLANs) in recent years, is leading to very dense deployments where high levels of channel contention may prevent to meet the increasing users' demands. To mitigate the negative effects of channel contention, the Target Wake Time (TWT) mechanism included in the IEEE 802.11ax amendment can have a significant role, as it provides an extremely simple but effective mechanism to schedule transmissions in time. Moreover, in addition to reduce the contention between stations, the use of TWT may also contribute to take full advantage of other novel mechanisms in the IEEE 802.11 universe, such as multiuser transmissions, multi-AP cooperation, spatial reuse and coexistence in high-density WLAN scenarios. Overall, we believe TWT may be a first step towards a practical collision-free and deterministic access in future WLANs.

\vspace{0.5cm}
{\bf \textit{Keywords}:} IEEE 802.11, WLANs, CSMA/CA, dense networks
\end{abstract}


\onehalfspacing


\section{Introduction} 

The 11ax Task Group \cite{11axTG} was created within the IEEE 802.11 working group in 2013 with the goal of enhancing the PHY and MAC layers so as to improve spectrum utilization efficiency in dense WLAN scenarios, improving users' performance accordingly, while ensuring efficient power stations' consumption. To do that, the IEEE 802.11ax amendment\footnote{It is expected that the IEEE 802.11ax amendment will be published in 2019.} \cite{IEEE80211axDraft,bellalta2016ieee} supports both Uplink (UL) and Downlink (DL) Multiuser (MU) transmissions based on Orthogonal Frequency Division Multiple Access (OFDMA) and Multiple Input Multiple Output (MU-MIMO), channel widths of up to 160 MHz, and spatial reuse solutions based on adapting the transmit power level and sensitivity threshold. On the other hand, to support low energy consumption, in addition to the default IEEE 802.11 power saving mechanisms, the IEEE 802.11ax amendment further develops the Target Wake Time (TWT) mechanism introduced in the IEEE 802.11ah amendment~\cite{khorov2015survey} aiming to provide a low-consumption mode for stations with low traffic requirements, and periodic data transmissions, such as those from Internet of Things (IoT) applications~\cite{adame2014ieee}.

Using TWT, each station negotiates awake periods with the Access Point (AP) to transmit and receive data packets, saving energy the rest of the time as the station remains in doze mode. In the one hand, this leads to a low energy consumption for the participating stations, but in the other hand, it also reduces the contention level significantly, even supporting a collision-free and deterministic operation when stations are distributed over different TWT sessions. Therefore, in scenarios with multiple heterogeneous traffic flows the use of TWT can further improve the spectrum utilization if a proper awake/sleep schedule is followed, that is, a schedule able to give to each flow the required amount of time and transmission opportunities. Note that the Point Coordination Function (PCF) of the IEEE 802.11 MAC is now obsolete and may be not included in later versions of the standards. In addition, it provides only a limited set of the features and benefits of TWT. Indeed, TWT provides a higher flexibility in how the transmission period can be used, as it allows both uplink and downlink traffic and does not limit transmissions only to those poll-based. Moreover, stations are not required to listen to all Beacons, significantly reducing energy consumption.

Following such a disruptive use of the TWT mechanism, we first overview the main features of the IEEE 802.11ax amendment, with the focus placed on its new MU capabilities. Then, we introduce the two TWT types of agreement, the individual and the broadcast, and we perform a performance assessment of the TWT management overheads and gains in terms of packet delay and channel utilization when scheduled access is enabled. There, it is shown that using TWT we can easily benefit from the MU and packet aggregation capabilities of IEEE 802.11ax as groups of stations can be scheduled to wake up at the same pre-defined instants of time. Finally, we discuss several other potential uses of TWT to improve quality of service provision, spatial reuse, coexistence issues and multi-AP coordination for next-generation WLANs. Overall, we envision the TWT mechanism as an step forward towards deterministic WLANs where service guarantees can be provided.


\section{Multiuser transmissions in IEEE 802.11ax}\label{sec:basic} 

Before entering into the details of the TWT mechanism, we briefly overview some of the main features of the IEEE 802.11ax multiuser capabilities, as we will refer to them later. Compared to IEEE 802.11ac, IEEE 802.11ax further extends WLANs multiuser capabilities by introducing Orthogonal Frequency Division Multiple Access (OFDMA) and uplink multiuser MIMO (MU-UL-MIMO) transmissions, while keeps the support for downlink multiuser MIMO (DL-MU-MIMO). In summary, multiuser transmissions in IEEE 802.11ax, in both the downlink and the uplink, can be performed through OFDMA, MU-MIMO, or a combination of them. A detailed description of IEEE 802.11ax MU operation can be found in \cite{bellalta2017ap}.

\begin{figure}[t]
\centering
\epsfig{file=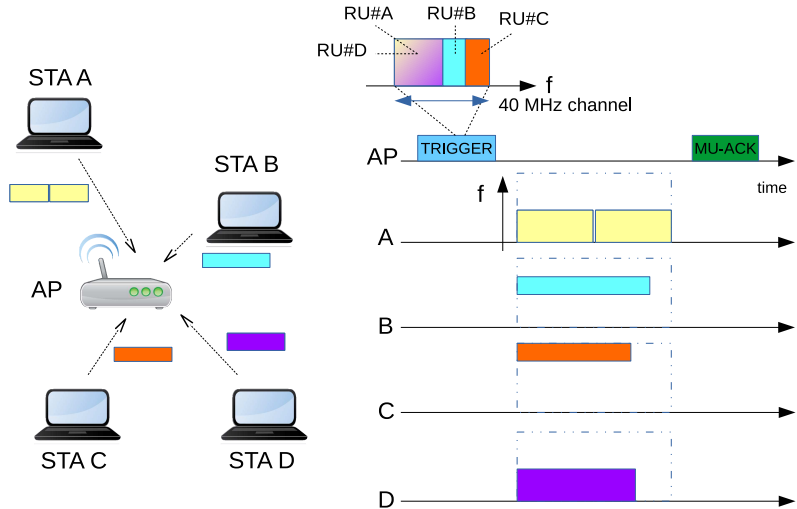,width=0.65\columnwidth,angle=-0}
\caption{Example of uplink MU transmission scheduled by the AP using OFDMA.}
\label{Fig:Figura_11axMU}
\end{figure}

Both DL and UL MU transmissions are scheduled by the Access Point (AP). In case of UL transmissions through Trigger frames reporting instructions to the involved stations. If OFDMA is used, the AP assigns to each station a set of subcarriers, called Resource Units (RUs), going from $26$--tones ($26$ subcarriers) to $2$x$996$--tones, depending on the channel width used: $20$, $40$, $80$ or $160$ MHz. Alternatively, for UL OFDMA transmissions, the AP can specify a common set of RUs allocated for random access to the contending stations. In order to use this common set of RUs allocated for random access, stations take into account the information sent by the AP periodically in the beacons, such as the OFDMA contention window (minimum and maximum values), so that each station is able to calculate its own OFDMA Backoff counter value.

Similarly, to enable MU-MIMO transmissions, the AP uses the Trigger frames to allocate subsets of spatial streams to each station, supporting up to $8$ simultaneous transmissions per RU, with the constraint that those RUs in which MU-MIMO transmissions are allowed must be equal or larger than 106 subcarriers. Figure~\ref{Fig:Figura_11axMU} shows an AP scheduling an MU-UL transmission using OFDMA and MU-MIMO over a 40 MHz channel involving stations A, B, C and D. The trigger frame contains the RUs and spatial stream allocation for all four stations (A and D share an RU of 242 subcarriers using MU-MIMO, and B and C have allocated different RUs of 106 subcarriers each), which is used by the selected stations to transmit their MAC Protocol Data Units (MPDUs) to the AP. Note that station A is able to aggregate two MPDUs while stations B, C and D only transmit one. Finally, the uplink transmissions are acknowledged with a single Multiuser ACK.

For an efficient resource allocation, the AP periodically requests to the stations to send both Channel Sounding and Buffer Status Reports (BSRs) indicating the channel conditions they observe, and the amount of buffered traffic they have, respectively. In addition, stations may report the bandwidth availability for each $20$ MHz channel in use by sending Bandwidth Query Reports (BQR), that is, the fractions of the channel width that are sensed as busy or idle by the stations.


\section{TWT Operation in IEEE 802.11ax}\label{twtax} 

With the TWT mechanism, stations can agree with the AP on a common wake scheduling, allowing them to wake up only when required, hence to minimize energy consumption and contention within the Basic Service Set (BSS), that is, the wireless network formed by the AP and the associated stations. Two are the key concepts to understand how TWT works. The TWT session or Session Period (SP) is the time period in which a station is awake to receive or send data. The TWT Agreement is the final arrangement between the AP and the station, reached after negotiation, to define the details of the TWT SP(s) the station will belong to, e.g., the time(s) the station has to wake up. Indeed, one TWT agreement allows the station to participate in multiple TWT SPs which wake up periodically. Note that the amendment does not specify any criteria for selecting stations for a TWT session or how to schedule them, thus it depends on the implementation. A TWT agreement may allow DL, UL or both types of transmissions, according to the negotiation and to further instructions that the AP can provide at the beginning of each TWT SP, e.g., through a Trigger frame. Finally, it is worth to mention that stations using TWT are allowed to wake up at any instant of time to initiate a new transmission using the Distributed Coordination Function (DCF).

The TWT mechanism in IEEE 802.11ax is based on the implementation in IEEE 802.11ah. In addition to the Individual TWT, it also includes the Broadcast TWT. It is expected that this new type of agreement will be able to improve the efficiency of the TWT operation and further leverage the new multiuser capabilities of IEEE 802.11ax.


\subsection{Individual TWT agreement}\label{sec:indiv}

To initiate a TWT session, first there is a negotiation phase in which the AP and the target station agree on a common set of parameters, among which the most relevant are:
\begin{itemize}
\item \textit{Target Wake Time (TWT)}: next time in microseconds at which the station participating in the TWT-based communication should wake up for the TWT session. 
\item \textit{TWT Wake interval}: the time interval between subsequent TWT sessions for the station; the value is higher than 0 when the TWT is periodic.
\item \textit{Minimum TWT wake duration}: minimum time duration a station shall stay awake since the starting time of the TWT session so as to be able to receive frames from the other station(s)\footnote{The minimum value is $256 \mu sec$}.
\item \textit{TWT Channel}: the channel a station can use temporarily as the primary one, similar to the Subchannel Selective Transmission of IEEE 802.11ah.
\item \textit{TWT Protection}: mechanism employed to protect a TWT session from transmissions of external stations, such as the RTS/CTS.
\end{itemize}

\begin{figure}[t!]
\centering
\subfloat[Individual TWT agreement: Non-trigger, Unannounced \& periodic packet exchange between AP and station $i$.]{\epsfig{file=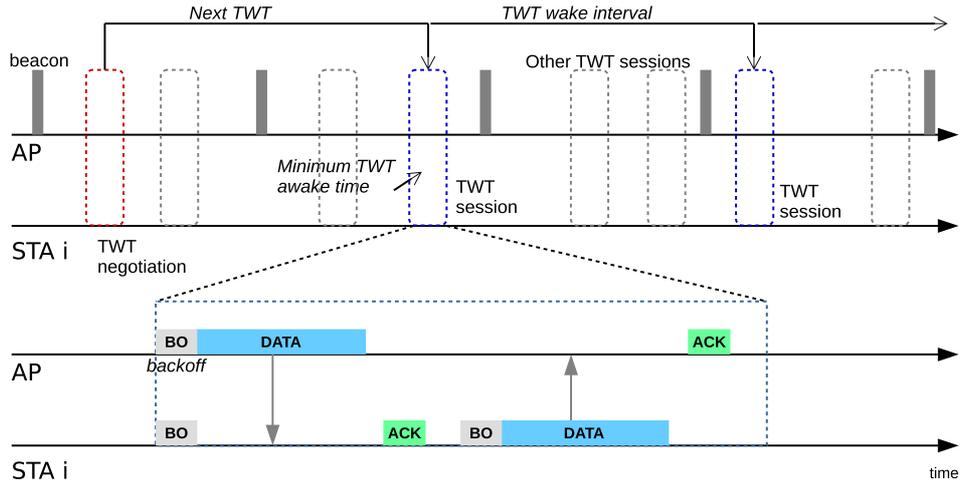,width=0.8\columnwidth,angle=-0}\label{Fig:Figura1_Individual}}\\
\subfloat[Broadcast TWT agreement: Trigger-enabled \& Unannounced packet exchange between AP and stations $i$ and $j$.]{\epsfig{file=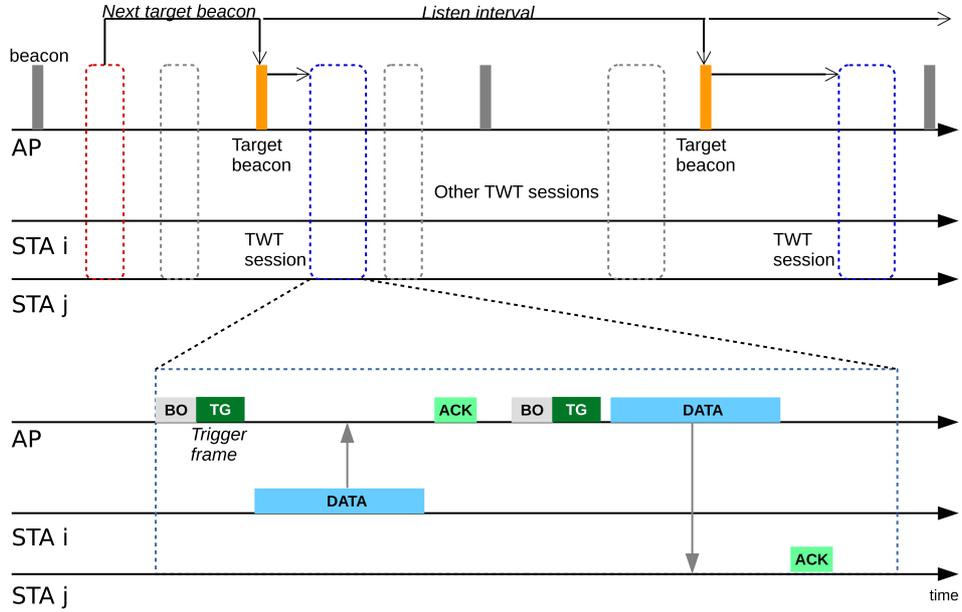,width=0.8\columnwidth,angle=-0}\label{Fig:Figura1_Broadcast}}
\caption{Example of the Individual and Broadcast TWT operation with SU transmissions.}
\label{Fig:Figura1}
\end{figure}

During the negotiation phase, also the following aspects are defined. First, the TWT agreement can be \textit{explicit} or \textit{implicit}: while the former requires specification of the TWT parameters before each new session, the latter allows to set periodic sessions by relying on the first set of parameters until a new set is received. In addition, two different operation modes inside a TWT session exist: 1) \textit{Trigger--enabled}, where the AP uses Trigger frames to schedule stations' transmissions, and 2) \textit{Non Trigger--enabled}, where the use of Trigger frames is not required, thus allowing each station to decide when to transmit autonomously inside the TWT session. Finally, when an AP sets up a TWT session with a station (or a set of stations, see below), it can be specified as \textit{Announced} or \textit{Unannounced}. The former implies that station(s) have to send messages to the AP to require buffered data. The latter allows the AP to deliver data without waiting any previous frame from the station(s), leveraging the fact that all stations from a TWT session must be awake when it starts. Once the TWT parameters are agreed, the stations can go to sleep until the next TWT session is reached. In practice, each station can establish up to 8 TWT agreements with the AP, as each agreement is identified by a 3-bit value. The TWT individual agreement is depicted in Figure~\ref{Fig:Figura1_Individual}.

Let us describe in more detail how the negotiation phase works. In IEEE 802.11ax, the \textit{Requesting} station is the one initiating the set up of a TWT session, whereas the \textit{Responding} station is the one accepting or rejecting the request. In principle the standard allows any station to have an agreement with any station, but in practice the description and definition of all the negotiation process implies the agreement between the AP and a station. For this reason and to avoid confusion, from now on we always consider that the AP specifies the schedule during the TWT sessions, whereas other cases are left for future consideration.

To create a new TWT session, the requesting station sends a TWT Request message to the responding station. In the TWT Request message it specifies the parameters for the TWT session, including the minimum TWT wake duration, TWT wake interval and TWT channel. Moreover, the request message type has to be one of the following:
\begin{itemize}
\item \textit{Suggest}: the set of parameters' values included in the request are those that the requesting station is willing to use, but it will consider accepting an alternative set.
\item \textit{Request}: the requesting station is willing to set a TWT agreement and lets the responding station specify the TWT parameters' set.
\item \textit{Demand}: the requesting station wants to set an agreement but will not accept a set of parameters different than the one in the request message.
\end{itemize}
On the other hand, the responding station can send a response of the following types:
\begin{itemize}
\item \textit{Accept}: the responding station accepts the request and the TWT agreement is set up with the parameters' values specified in the response message.
\item \textit{Alternate}: the responding station proposes an alternative set of parameters' values.
\item \textit{Dictate}: the responding station demands another set of parameters with no possibility for further negotiating them.
\item \textit{Reject}: the TWT session is not accepted.
\end{itemize}

\begin{figure}[t]
\centering
\epsfig{file=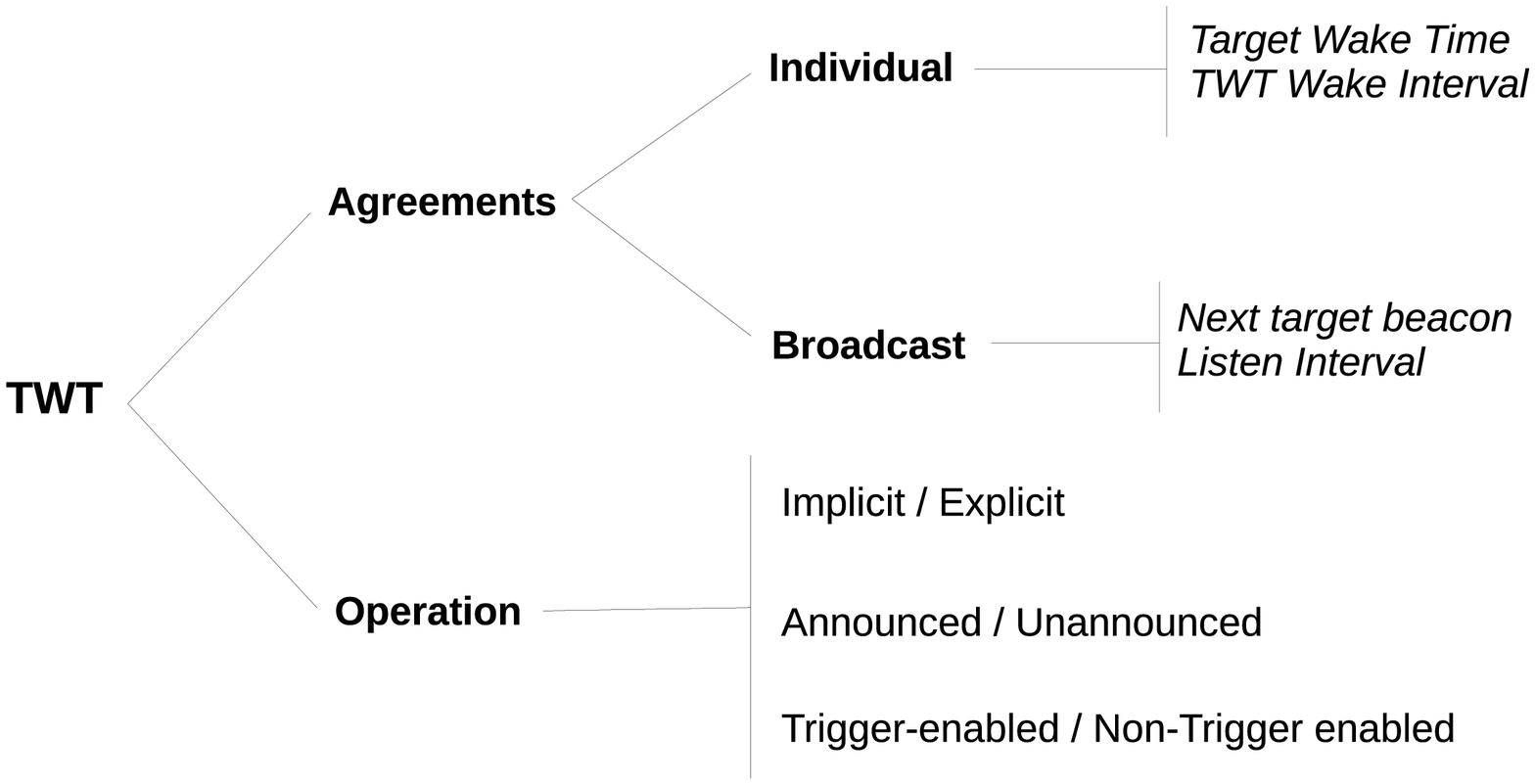,width=0.65\columnwidth,angle=-0}
\vspace{0.5cm}
\epsfig{file=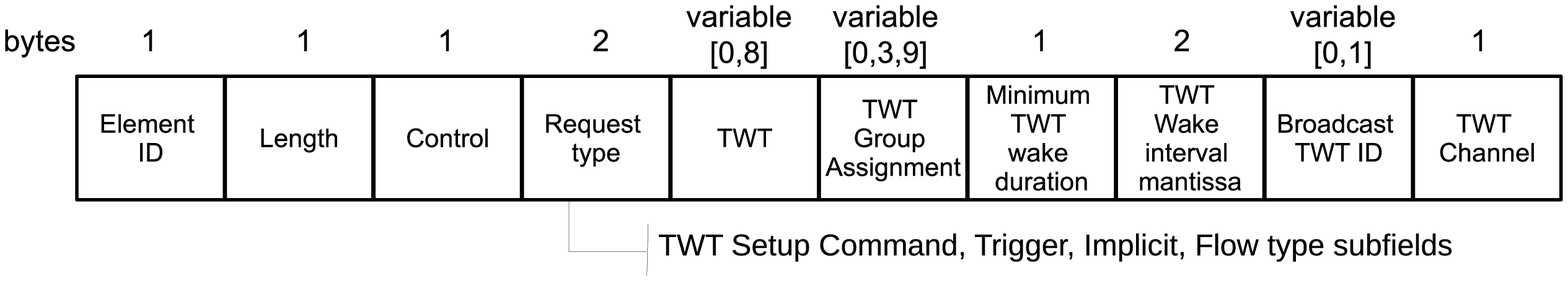,width=0.75\columnwidth,angle=-0}
\caption{Overview of TWT modes and key parameters, and TWT request/response frame}
\label{Fig:TWT_IDElement}
\end{figure}

The second and third option would require (at least) another pair of request-response messages to conclude the agreement negotiation phase. When the final response is of type \textit{Accept}, a TWT agreement has been setup. From this moment on, the requesting station may go to sleep until the next TWT session starts. All the details of the TWT agreement are carried within a TWT Element, as shown in Figure \ref{Fig:TWT_IDElement}, which can be included in any frame exchanged between the AP and a station for the association process. The \textit{Request type} field contains some important subfields. The \textit{TWT Setup Command} field specifies the type of message (e.g., request, suggest, accept). In addition, the fields \textit{Trigger} and \textit{Implicit} are used to specify the operation mode, and the field \textit{Flow type} specifies if it is Announced or Unannounced.

The AP can have multiple TWT agreements, each one with a different station, but some of them may overlap in time. In this case, the stations may be scheduled by the AP for simultaneous transmissions, or have to contend for the medium through random access, as explained in Section \ref{sec:basic}. In addition, the TWT \textit{grouping} mechanism allows the AP to specify a TDMA-like scheduling from the start time of the common TWT session by providing a transmission time for each group and for each station within the group. The amendment does not strictly define how the AP should group stations or schedule transmissions, thus these options open many interesting research issues, and offer space for investigating the most suitable scheduling approach combining TWT with the new multiuser capabilities of IEEE 802.11ax. A deeper discussion is provided in Section \ref{sec:advanced}. An overview of TWT options is shown in Figure \ref{Fig:TWT_IDElement}.

\subsection{Broadcast TWT agreement}\label{sec:broadc}

The \textit{Broadcast TWT} operation allows an AP to set up a shared TWT session for a group of stations, and specify periodically the TWT parameters set within Beacon frames. The stations of a TWT Broadcast agreement are required to wake up to receive only the Beacons containing instructions for the TWT Broadcast sessions they belong to. Note that the AP may advertise existing TWT Broadcast agreements so that stations may ask membership to existing TWT sessions, or send requests to create new ones. The TWT Broadcast operation is illustrated in Figure \ref{Fig:Figura1_Broadcast}.

In order to request the participation in a Broadcast TWT agreement, a station has to send a TWT Request to the AP. Such a request can be sent also in response to a participation request solicited by the AP to all the associated stations that support TWT. Similarly to the Individual case, in the negotiation phase the station can request, suggest or demand the set of TWT parameters of the Broadcast TWT session, and the AP can accept or reject the request, or propose an alternative setting. In all cases, the TWT parameters are decided by the AP.

During the setup phase, the station may also negotiate other two fundamental parameters:
\begin{itemize}
\item \textit{Next target beacon}: the next transmission time of a Beacon including TWT information relevant for the station, i.e., related to the Broadcast TWT session the station belongs to. 
\item \textit{Listen Interval}: interval between subsequent beacons carrying TWT information relevant for the station.
\end{itemize}

Then, the station goes to doze state and wakes up at the time at which the next relevant Beacon is scheduled. These Beacons carry the necessary information about the Broadcast TWT session that allow the involved stations to follow the session schedule. In addition, the AP may also broadcast any update on the TWT parameters's set of the session so that all the involved stations can properly update it.

As in the Individual agreement, the Broadcast TWT session can be either \textit{Trigger--enabled} or not, and it can be implemented as \textit{Announced} or \textit{Unannounced}, working the same way as explained in the previous section.


\subsection{Management overheads} 

In both individual and broadcast TWT agreements the setup phase of a TWT session requires the exchange of several messages. As explained before, a \textit{Request} and \textit{Accept} is the minimum set of messages, but multiple pairs of \textit{Suggest/Demand} and \textit{Alternate/Dictate} may be exchanged before reaching the final agreement. In order to get some insights on the extra overheads TWT management may cause, as an example, we overview the number of control messages exchanged for a single TWT agreement during $1$ hour for the following four TWT operation modes (based on the explanation in Sections \ref{sec:indiv} and \ref{sec:broadc}):
\begin{itemize}
\item \textit{Individual Periodic (IP)}: Individual implicit agreement, that is, the station needs only to know when the first TWT session starts and the TWT interval, from which it can calculate all the subsequent TWT values.
\item \textit{Individual Aperiodic (IA)}: Individual explicit agreement, that is, the station needs to know the next TWT value before any TWT session. At least a message directed to each station must be send to inform it about next TWT session.
\item \textit{Broadcast Periodic (BP)}: Broadcast implicit agreement, that is, the station needs only to know the first TWT value and the TWT interval, from which it can calculate all the subsequent TWT values.
\item \textit{Broadcast Aperiodic (BA)}: Broadcast explicit agreement, that is, the station needs to know the next TWT value before any TWT session. A single message containing the identifier of the TWT session can be used to reach all the stations, and inform them about next TWT session.
\end{itemize}

\begin{table}[t!]

	\centering
	\small
  \begin{tabular}{ p{1.5cm} | p{1.5cm} | p{1.5cm} | p{2cm} }
 \textbf{Number of stations ($N$)} & \textbf{Number of updates / hour ($k$)} & \textbf{TWT mode} & \textbf{Total number of control messages in 1 hour} \\ \hline
 \multirow{4}{*}{10} & \multirow{4}{*}{10} & IP & 20 \\ 
   &  & IA & 120 \\ 
   &  & BP & 20 \\ 
   &  & BA & 30 \\ \hline 
  \multirow{4}{*}{10} & \multirow{4}{*}{100} & IP & 20  \\ 
   &  & IA & 1020  \\ 
   &  & BP & 20 \\ 
   &  & BA & 30  \\ \hline 
 \multirow{4}{*}{100} & \multirow{4}{*}{10} & IP & 200 \\ 
   &  & IA & 1200  \\ 
   &  & BP & 200  \\ 
   &  & BA & 210  \\ \hline 
  \multirow{4}{*}{100} & \multirow{4}{*}{100} & IP & 200  \\ 
   &  & IA & 10200  \\ 
   &  & BP & 200  \\ 
   &  & BA & 300  \\ \hline 
  \end{tabular}

  \caption{Example of TWT overheads. The values of $10$ and $100$ updates/hour mean that updates are sent every $6$ minutes and $36$ seconds respectively.  }
  \label{Tbl:Overheads}
\end{table}


For simplicity and without loss of generalization, we assume that the minimum sequence of two messages is used for both individual and broadcast TWT arrangements. Similarly, for updating parameters of an existing TWT agreement, we assume that only one message is needed\footnote{We focus on the general operation mode and do not consider other sequences that may be used only in some cases and may require a variable number of messages.}. Finally, we assume that the AP only updates Aperiodic TWT agreements (IA and BA). Thus, the number of updates $k$ only affects the Aperiodic cases, as in the Periodic cases (IP and BP) the station can calculate the next TWT based on the information received during the negotiation.

For the calculations, we vary both the number of stations $N$ that request to initiate a TWT session, and the number of updates the AP sends to the stations in $1$ hour ($k$). Table \ref{Tbl:Overheads} shows the amount of messages required for different $N$ and $k$ values. Clearly, the values of the Periodic cases are only affected by the number of stations, each one requiring only two messages for the initial negotiation. Note that broadcast sessions require the exchange of less packets as updates to multiple stations can be sent in a single beacon. Overall, in both cases, the TWT management overheads are significantly low even for a large number of stations and frequent updates. Therefore, TWT management overhead only implies a small fraction of channel airtime even if management packets are transmitted at the lowest available transmission rate. Only in the case of IA with $N=k=100$ more than $3$ packets are exchanged per second, which assuming a duration of $2$ ms per exchange, roughly represent about the $6$~\% of the airtime.


\subsection{Performance Assessment of TWT}\label{perf2} 

\begin{figure}[t!]
	\centering
	\small
	\begin{tabular}{c|c}
	\textbf{Parameter} & \textbf{Value }\\
	\hline \hline
	Transmission Power & 16 dBm \\
	Path-loss Model & TMB 5GHz indoor path-loss model \cite{TMBpathloss}\\
	Channel Width & 20 MHz \\	
	\hline
	Traffic model & Poisson Arrivals \\
	Traffic load per station & 1,2,4,6 and 8 Mbps \\	
	Buffer size & 500 packets \\
	\hline	
	Multi-user capabilities & MU-MIMO (up to 8 users) \\
	Max A-MPDU size & $64$ packets \\
	MPDU size & $12000$ bits \\
	Single-user MIMO streams & up to $2$ \\
	Channel and buffer sounding overheads & negligible \\
	\hline
	CW$_{\min}$ and CW$_{\max}$  & $15$ and $511$ slots \\
	Channel access mode & RTS/CTS, MU-RTS/CTS \\
	\hline
	TWT mode & Trigger-enabled individual TWT agreements \\
	TWT signalling overheads & negligible \\
	\hline
	\end{tabular}

	\vspace{0.5cm}

	\epsfig{file=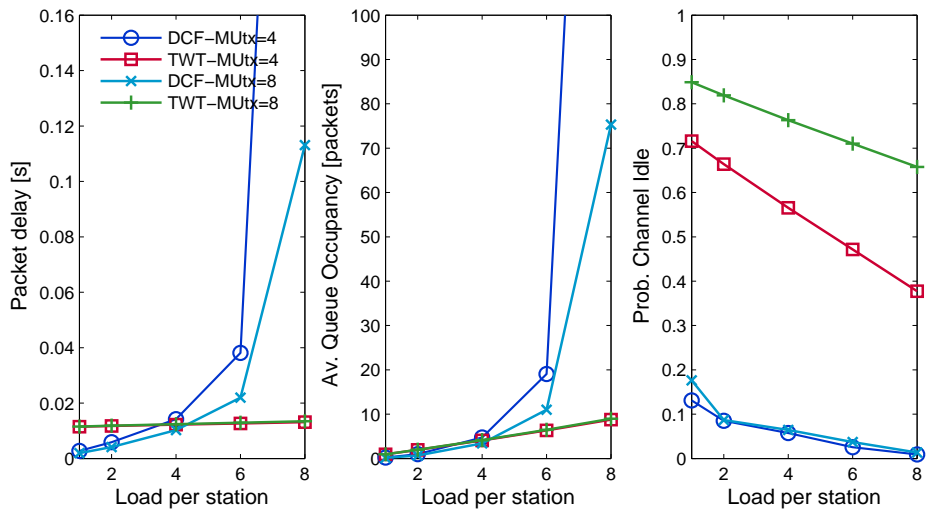,width=0.9\columnwidth,angle=-0}

	\caption{Simulation IEEE 802.11ax parameters and results. DCF vs TWT: delay, queue occupancy and fraction of time the channel is idle. In the legend, 'MUtx' indicates the maximum number of simultaneous (MU) transmissions supported, which in the case of TWT, corresponds with the number of stations in a TWT session too.}
	\label{Fig:Fig_Result2}		
\end{figure}

To assess the performance gains of TWT we consider a single WLAN that consists of an AP and $16$ stations. The stations are randomly placed in a 20x20m$^2$ area with the AP in its center. All stations are inside the coverage area of the AP. Only uplink traffic (i.e., from the stations to the AP) is considered. Figure~\ref{Fig:Fig_Result2} details the parameters used for the simulations. When TWT is considered, the $16$ stations are randomly placed in $2$ and $4$ different TWT sessions, with $8$ and $4$ stations each TWT session, respectively. Different TWT sessions do not overlap in time, but all of them activate periodically every $20$ milliseconds. All stations in the same TWT session are scheduled to transmit simultaneously using the MU capabilities. The number of packets transmitted by each station depends on the buffer occupancy. Packet aggregation is enabled, with a maximum Aggregate MAC Protocol Data Unit (AMPDU) size of $64$ packets. The simulator used to obtain the results is written in C++ extending the work done in \cite{bellalta2017ap}.

Figure \ref{Fig:Fig_Result2} shows the average packet delay, average buffer occupancy, and the fraction of time the channel is idle, when either DCF or TWT is used for different traffic loads. Note that the average packet delay when TWT is used is almost constant regardless the traffic load, and corresponds approximatively to half the TWT wake interval. These results illustrate the benefits and drawbacks of using TWT to schedule traffic flows in time. While for low traffic loads ($\leq~4$ Mbps) DCF results in a lower packet delay than TWT, the opposite is observed for higher traffic loads. This is because TWT enables more efficient MU transmissions and packet aggregation. Since using TWT, the AP is in charge of scheduling all UL transmissions by sending trigger frames at the beginning of a TWT session, all transmissions in the network can effectively be MU. Moreover, they can aggregate a large amount of packets as the inactivity time between consecutive TWT sessions allows to fill up the buffers of all stations. The more efficient use of the channel resources is also observed in the lower queue occupancy values for high traffic loads, but it is especially visualized in the fraction of the time the channel remains idle, which in case of TWT is far higher than using DCF. Indeed, that amount of idle time could be effectively used for the AP and other non-TWT stations to transmit their data, thus increasing the network capacity without affecting the performance of those stations using TWT. 


\section{Advanced TWT operation}\label{sec:advanced} 

The TWT mechanism was originally designed for reducing the energy consumption. However, it opens also the door to fully maximize the IEEE 802.11 MU capabilities by scheduling in time both MU-DL and MU-UL transmissions, for collecting information from the stations such as the channel sounding and buffers's occupancy in predefined periods, and to support other not yet implemented features such as full-duplex communication. In addition, in dense scenarios, several WLANs may use TWT to agree on non-overlapping schedules to further improve Overlapping BSSs (OBSS) coexistence, among others potential uses.


\subsection{Multiuser TWT operation}\label{sec:mutwt}

By assigning multiple stations to the same TWT session and awakening them at the same time, efficient scheduling of MU transmissions can be performed, using either MIMO or OFDMA capabilities, or both, while at the same time reducing the chances of collisions between the AP and stations, and so its negative effects \cite{bellalta2017ap}. Moreover, the periodic scheduling of MU transmissions may help to guarantee enough buffered traffic at the AP and at the stations to further exploit the use of packet aggregation.

This can be done for stations using the Trigger-enabled TWT mode, with either Individual or Broadcast agreements. Then, in each TWT session, after the time all stations wake up, the AP can send further instructions using trigger frames to enable MU transmissions from the stations, or to send data to them. Similarly, channel and buffer sounding (i.e., CSIs and BSRs), as well as spectrum occupancy observations (i.e., BQR), can also be scheduled periodically.


\subsection{Parallelization of TWT sessions}\label{sec:partwt}

In addition to simply using TWT to improve the scheduling of MU transmissions from the same TWT session, the other way around is to use MU transmissions to parallelize several TWT sessions that overlap in time (periodically, or sporadically) by simply partitioning the available resources between them. Although the concept is similar to the previous case, conceptually adds a new degree of freedom to easily support miscellaneous periodicities for TWT sessions without requiring to reschedule all active ones every time a TWT agreement is updated, or a new one is set-up, thus reducing also extra control overheads.


\subsection{Quality of Service}\label{sec:mutwt}

The use of TWT can enable different quality of service levels by simply scheduling TWT sessions with different periods, giving more transmission opportunities to stations with more, and/or high priority traffic. However, a clear open challenge is how to map the different Access Categories (AC) in EDCA to the different TWT periodicities. The most simple approach is to allocate TWT sessions in which only transmissions from certain ACs are allowed, thus instead of setting TWT sessions per station they would be set by the pair AC-station. Although conceptually very different, the scheduled access defined in CSMA with Enhanced Collision Avoidance (CSMA/ECA) for multiple ACs \cite{sanabria2015traffic} can provide inspiration to dynamically schedule TWT sessions supporting different priorities.

Nevertheless, a more interesting aspect is to simply perform per AC-station pair flow agreements when the traffic load is known or it can be estimated, such as for the case of video flows, thus reserving enough time and transmission opportunities to operate in a pure contention-free mode. For best-effort and sporadic traffic, a certain amount of time can also be allocated to them (i.e., a specific TWT session) to avoid flow starvation. Note however, in case that per-flow/connection TWT sessions are created, TWT overheads may increase significantly depending on the flow-size distribution.


\subsection{Coordination of multiple OBSSs}\label{sec:combineOBSS}

Another context where TWT can play a crucial role is to reduce the contention in presence of OBSS. Overlapping wireless networks may experience very high variability in the number of associated stations and traffic patterns, thus creating very complex, and difficult to predict in advance, interference patterns among them. Here, we briefly discuss how TWT can be used to improve channel contention between OBSSs through APs' coordination:

\begin{figure}[t]
\centering
\subfloat[Network sketch]{\epsfig{file=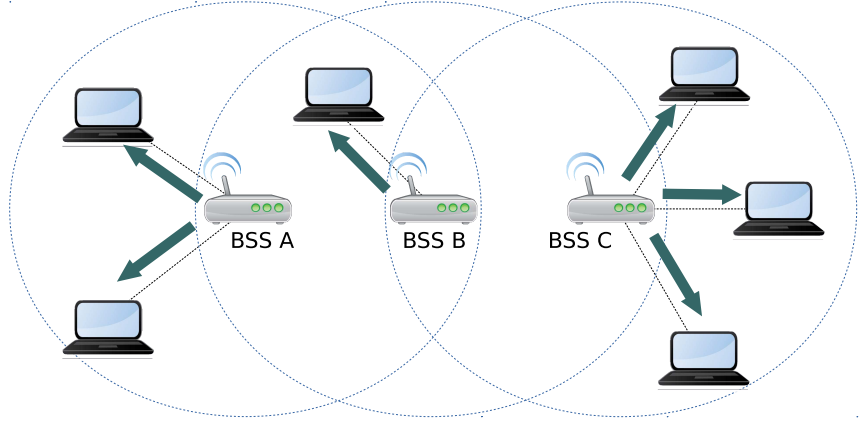,width=0.7\columnwidth,angle=-0}}\\
\subfloat[Temporal evolution]{\epsfig{file=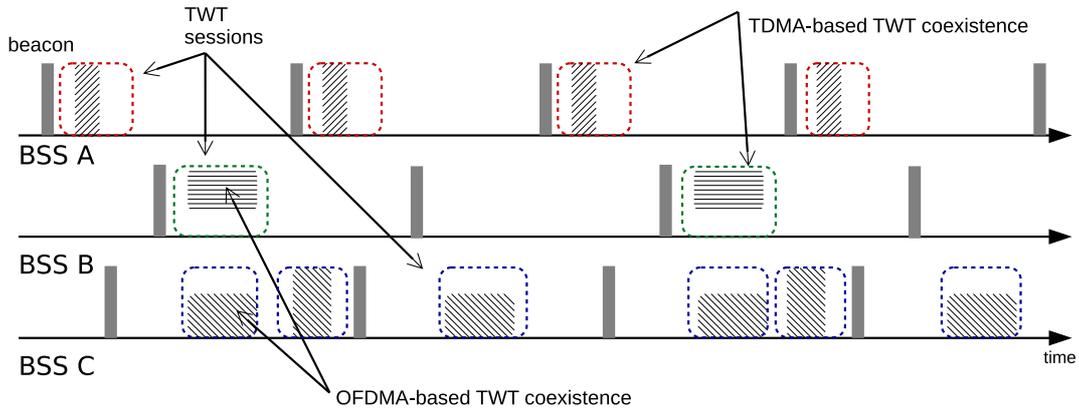,width=0.9\columnwidth,angle=-0}}
\caption{Three overlapping WLANs. BSS A and BSS B share the medium following a TDMA approach, while BSS B and BSS C avoid contention by using different OFDMA RUs. Notice that BSS C is able to use all RUs in those periods in which BSS B remains silent due to its TDMA schedule with A.}
\label{Fig:Figura_OBSS}
\end{figure}

\begin{itemize}
\item \textbf{Joint TWT scheduling}: Cooperation among APs in densely deployed IEEE 802.11 WLANs is fundamental for ensuring fair and resource-efficient coexistence \cite{kosek2017coexistence}. If each AP communicates the desired period(s) of time for its own TWT session(s) to a central entity, or directly to the other APs, a kind of TDMA scheduling for the OBSSs can be found. Moreover,  the temporal scheduling can be enhanced by considering also sharing resources within the OBSS at the RU level through OFDMA, which would allow a much more dynamic and fine grained solution. The combination of both TDMA and OFDMA will not only reduce the time a station is awake, but can also significantly minimize the contention level within and between OBSSs, a key aspect for future dense wireless scenarios. To illustrate the described cases, Figure \ref{Fig:Figura_OBSS} shows an example with three BSSs, where BSS A and B share the spectrum using a TDMA approach, while BSS B and C use OFDMA.
\item \textbf{Cooperative transmission techniques}: Moreover, in addition to sharing time or frequency resources between OBSSs, cooperative techniques such as Network MIMO \cite{zhang2013nemox} and Interference Alignment \cite{dovelos2017breaking} can benefit from the explicit synchronization introduced by TWT. Using those techniques, different APs may cooperate to transmit data simultaneously by leveraging the spatial dimension. However, to enable such techniques, further exchange of timely information such as channel state information between APs is required, adding extra complexity to the network design such as requiring very fast backhaul.
\item \textbf{Spatial reuse improvement}: Spatial reuse is a key aspect in future high-density WLANs in order to improve the area throughput \cite{selinis2016evaluation,wilhelmi2018potential}. By reducing the uncertainty with respect to the number of stations in a given area that may transmit in a given period of time, techniques to improve the spatial reuse by adapting the transmit power, tuning the Clear Channel Assessment (CCA) threshold, and selecting the channel to operate can be further improved. For instance, avoiding potential hidden and exposed nodes allows for a much more aggressive transmit power and CCA adaptation to reduce exposed node and starvation aspects in dense WLANs, which on turn, would allow to support a higher number of parallel TWT sessions from different BSSs, hence increasing the area throughput. The challenge here is to be able to identify which are the potential hidden and exposed nodes for each BSS, and find suitable schedules to avoid them. Although not considering TWT, a similar approach is developed in \cite{khorov2016joint} where TDMA schedules and DSC adaptation are used to improve the throughput in dense WLAN deployments.

\end{itemize}

How to implement OBSS coordination is outside the scope of this work. However, for completeness, it is worth to mention the current trend for enterprise WLANs to centralize all management and control functions in a single controller with a global vision of the network \cite{gallo2018cadwan}. We are convinced the aforementioned TWT use-cases can be easily implemented on top of those network management solutions.


\section{Conclusions}\label{concl} 

This paper has proposed some new disruptive uses of TWT to improve the operation of next-generation WLANs by explicitly reaching a collision-free operation. The TWT operation allows stations to be awake only during certain periods of time, thus significantly improving channel contention if a proper schedule is found. We have shown that significant performance gains in throughput can be achieved when combined with the MU capabilities of IEEE 802.11ax. Moreover, other currently active areas in WLAN research can greatly benefit from the use of TWT to propose effective solutions, such as improving spatial reuse and coexistence techniques in dense WLAN scenarios thanks to the uncertainty reduction TWT is able to provide.

Nevertheless, there are still many open aspects in TWT operation that require further research. For example, the TWT operation in mixed scenarios with both TWT and non-TWT stations may require the design of new prioritization and protection strategies for TWT exchanges, the design of optimal strategies to group and assign stations to proper (broadcast) TWT sessions is still completely open, as well as the performance evaluation of the impact of signalling overheads, heterogeneous stations and traffic loads in dense WLAN scenarios. 


	\section*{Acknowledgment}
	
	This  work  has  been  partially  supported by a Gift from the Cisco University Research Program (CG\#890107, Towards Deterministic Channel Access in High-Density WLANs) Fund, a corporate advised fund of Silicon Valley Community Foundation, and by  the  Spanish Ministry of Economy and Competitiveness under the Maria de Maeztu  Units  of  Excellence  Programme  (MDM-2015-0502), Catalan Government (SGR-2017-1188). We would also like to acknowledge the constructive comments received from the anonymous reviewers. 

\bibliographystyle{unsrt}
\bibliography{Bib_Madda2}

\begin{thebibliography}{10}

\bibitem{11axTG}
IEEE~802.11 WLANs.
\newblock {The Working Group for WLAN Standards}.
\newblock 2018.

\bibitem{IEEE80211axDraft}
IEEE.
\newblock {Proposed TGax draft specification. doc.: IEEE P802.11ax/D2.0,
  October 2017}.
\newblock Technical report, IEEE, 2017.

\bibitem{bellalta2016ieee}
Boris Bellalta.
\newblock {IEEE 802.11ax: High-efficiency WLANs}.
\newblock {\em IEEE Wireless Communications}, 23(1):38--46, 2016.

\bibitem{khorov2015survey}
Evgeny Khorov, Andrey Lyakhov, Alexander Krotov, and Andrey Guschin.
\newblock {A survey on IEEE 802.11 ah: An enabling networking technology for
  smart cities}.
\newblock {\em Computer Communications}, 58:53--69, 2015.

\bibitem{adame2014ieee}
Toni Adame, Albert Bel, Boris Bellalta, Jaume Barcelo, and Miquel Oliver.
\newblock {IEEE 802.11ah: the WiFi approach for M2M communications}.
\newblock {\em IEEE Wireless Communications}, 21(6):144--152, 2014.

\bibitem{bellalta2017ap}
Boris Bellalta and Katarzyna Kosek-Szott.
\newblock {AP-initiated multi-user transmissions in IEEE 802.11 ax WLANs}.
\newblock {\em Ad Hoc Networks}, 85:145--159, 2019.

\bibitem{TMBpathloss}
Toni Adame, Marc Carrascosa, and Boris Bellalta.
\newblock {The TMB path loss model for 5 GHz indoor WiFi scenarios: On the
  empirical relationship between RSSI, MCS, and spatial streams}.
\newblock {\em arXiv:1812.00667 [cs.NI]}, 2018.

\bibitem{sanabria2015traffic}
Luis Sanabria-Russo and Boris Bellalta.
\newblock {Traffic Differentiation in Dense Collision-free WLANs using
  CSMA/ECA}.
\newblock {\em Ad-Hoc Networks}, 2018.

\bibitem{kosek2017coexistence}
Katarzyna Kosek-Szott, Janusz Gozdecki, Krzysztof Loziak, Marek Natkaniec,
  Lukasz Prasnal, Szymon Szott, and Michal Wagrowski.
\newblock {Coexistence Issues in Future WiFi Networks}.
\newblock {\em IEEE Network}, 31(4):86--95, 2017.

\bibitem{zhang2013nemox}
Xinyu Zhang, Karthikeyan Sundaresan, Mohammad A~Amir Khojastepour, Sampath
  Rangarajan, and Kang~G Shin.
\newblock {NEMOx: Scalable network MIMO for wireless networks}.
\newblock In {\em Proceedings of the 19th annual international conference on
  Mobile computing \& networking}, pages 453--464. ACM, 2013.

\bibitem{dovelos2017breaking}
Konstantinos Dovelos and Boris Bellalta.
\newblock {Breaking the Interference Barrier in Dense Wireless Networks with
  Interference Alignment}.
\newblock {\em IEEE ICC 2018}, 2018.

\bibitem{selinis2016evaluation}
Ioannis Selinis, Marcin Filo, Seiamak Vahid, Jonathan Rodriguez, and Rahim
  Tafazolli.
\newblock {Evaluation of the DSC algorithm and the BSS color scheme in dense
  cellular-like IEEE 802.11 ax deployments}.
\newblock In {\em Personal, Indoor, and Mobile Radio Communications (PIMRC),
  2016 IEEE 27th Annual International Symposium on}, pages 1--7. IEEE, 2016.

\bibitem{wilhelmi2018potential}
Francesc Wilhelmi, Sergio Barrachina-Mu{\~n}oz, Boris Bellalta, Cristina Cano,
  Anders Jonsson, and Gergely Neu.
\newblock {Potential and Pitfalls of Multi-Armed Bandits for Decentralized
  Spatial Reuse in WLANs}.
\newblock {\em Journal of Network and Computer Applications}, 2018.

\bibitem{khorov2016joint}
Evgeny Khorov, Anton Kiryanov, Alexander Krotov, Pierluigi Gallo, Domenico
  Garlisi, and Ilenia Tinnirello.
\newblock {Joint Usage of Dynamic Sensitivity Control and Time Division
  Multiple Access in Dense 802.11 ax Networks}.
\newblock In {\em International Workshop on Multiple Access Communications},
  pages 57--71. Springer, 2016.

\bibitem{gallo2018cadwan}
Pierluigi Gallo, Katarzyna Kosek-Szott, Szymon Szott, and Ilenia Tinnirello.
\newblock {CADWAN: A Control Architecture for Dense WiFi Access Networks}.
\newblock {\em IEEE Communications Magazine}, 56(1):194--201, 2018.

\end{thebibliography}

	\section*{Biographies}

	\small
	\noindent
	MADDALENA NURCHIS is a postdoctoral researcher and Visiting Professor at the University of Pompeu Fabra (UPF) in Barcelona. My current research activity is focused on resource allocation and performance evaluation of dense WLAN deployments and coexistence of LTE and WiFi in Unlicensed Spectrum.\\~\\
	BORIS BELLALTA is an Associate Professor in the Department of Information and Communication Technologies (DTIC) at Universitat Pompeu Fabra (UPF). His research interests are in the area of wireless networks, with emphasis on the design and performance evaluation of new architectures and protocols. 

\end{document}